\documentstyle[12pt]{article}
\input epsf

\begin{document}

\begin{titlepage}

\begin{flushright}
UCLA-00-TEP-15\\
UM-TH-00-07\\
March 2000
\end{flushright}
\vspace{2.0cm}

\begin{center}
\large\bf
{\LARGE\bf Exact non-factorizable ${\cal O}(\alpha_s g^2)$ 
           two-loop contribution to $Z \rightarrow b\bar{b}$}\\[2cm]
\rm
{Adrian Ghinculov$^a$ and York-Peng Yao$^b$}\\[.5cm]

{\em $^a$Department of Physics and Astronomy, UCLA,}\\
      {\em Los Angeles, California 90095-1547, USA}\\[.1cm]
{\em $^b$Randall Laboratory of Physics, University of Michigan,}\\
      {\em Ann Arbor, Michigan 48109-1120, USA}\\[3.0cm]
      
\end{center}
\normalsize

\begin{abstract}
For $Z \rightarrow b\bar{b}$, we calculate all the two-loop top 
dependent Feynman graphs, which have mixed QCD and electroweak 
contributions that are not factorizable. 
For evaluating the graphs, without resorting to a mass expansion,
we apply a two-loop extension
of the one-loop Passarino-Veltman reduction.
This is an analytic-numerical method, 
which first converts all diagrams into a set of ten standard 
scalar functions, and then integrates them numerically over the remaining
Feynman parameters, with rapid convergence and high accuracy.
We discuss the treatment of infrared singularities within our methods.
We do not resort to unitarity cuts of two-point functions for calculating 
decay rates; these are useful only to 
obtain an inclusive rate. For this reason,
experimental cuts and the experimental infrared energy resolution can be
implemented in our calculation, once the corresponding one-loop 
gluon Bremsstrahlung process is added to this calculation.  
\end{abstract}

\vspace{3cm}

\end{titlepage}


\title{Exact non-factorizable ${\cal O}(\alpha_s g^2)$ two-loop
           contribution to $Z \rightarrow b\bar{b}$}

\author{Adrian Ghinculov$^a$ and York-Peng Yao$^b$}

\date{{\em $^a$Department of Physics and Astronomy, UCLA,}\\
      {\em Los Angeles, California 90095-1547, USA}\\
      {\em $^b$Randall Laboratory of Physics, University of Michigan,}\\
      {\em Ann Arbor, Michigan 48109-1120, USA}}

\maketitle

\begin{abstract}
For $Z \rightarrow b\bar{b}$, we calculate all the two-loop top 
dependent Feynman graphs, which have mixed QCD and electroweak 
contributions that are not factorizable. 
For evaluating the graphs, without resorting to a mass expansion,
we apply a two-loop extension
of the one-loop Passarino-Veltman reduction.
This is an analytic-numerical method, 
which first converts all diagrams into a set of ten standard 
scalar functions, and then integrates them numerically over the remaining
Feynman parameters, with rapid convergence and high accuracy.
We discuss the treatment of infrared singularities within our methods.
We do not resort to unitarity cuts of two-point functionsfor calculating 
decay rates; these are useful only to 
obtain an inclusive rate. For this reason,
experimental cuts and the experimental infrared energy resolution can be
implemented in our calculation, once the corresponding one-loop 
gluon Bremsstrahlung process is added to this calculation.  
\end{abstract}


High energy experimental data have reached such an accuracy that higher
loop effects have to be accounted for.  It is increasingly clear that 
new physics most likely lies beyond the rubric of tree or even 
one-loop graphs of the standard model.  

To asses these higher order effects, in quite a few processes 
where one can identify a single large scale, a very powerful 
technique is asymptotic momentum or large mass expansion
\cite{massexpansion,topexpansion,degrassi}.  This 
has been applied successfully in several instances, where either 
the external momenta are small compared to the internal masses,
or vice versa.  However, in many examples, the external momenta 
are comparable to the internal scales, and therefore an expansion 
in their ratios does not apply a priori, 
or at least is not economical for convergence.

There is a further complication which must be addressed.  Massless 
particles and particles with very small masses are with us.   
Partly to avoid potential infrared and mass singularities, many 
authors calculate inclusive rates, where such issues are 
by-passed.  On the other hand, in a typical experimental setup one 
needs to make kinematical cuts among other things, and therefore 
exclusive processes must be considered.  It would help if 
a theoretical calculation can accommodate this.

In a previous article \cite{2loopgeneral} we gave a general 
framework to calculate any two-loop amplitude
with non-trivial interactions.  Our approach fully respects the 
mass structure and the kinematics of the physical process being 
investigated.  The
outcome is a standard set of ten functions and their derivatives, 
arising from tensorial decomposition. Numerical integration over 
the remaining Feynman parameters is then used for obtaining final results.

In this article we apply this general program 
to an ${\cal O}(\alpha_s g^2)$ calculation of the process 
$Z \to b \bar b$.
So far, this process was calculated by mass expansion
methods in refs. \cite{topexpansion}, 
with the gluon Bremsstrahlung process integrated over the whole phase space.
The calculations already available, being performed at higher order in 
the mass expansion, show that the expansion methods work 
well in this particular
process. For this reason, it is to be expected that the numerical output of
the calculation described
in this paper will not result in a sizeable difference from the existing
results of ref. \cite{topexpansion}. 
Our emphasis here is more on an illustration, based on an important
physical process, of how
our two-loop methods work, because the whole machinery deployed covers 
almost all typical situations, including particles with various masses, two-
and three-point functions, and treatment of infrared singularities. 

A complete two-loop calculation of the exclusive $Z \to b \bar b$ decay
consists of two technically different parts, namely the pure two-loop part,
and the evaluation of the one-loop Bremsstrahlung process. 
In this paper we treat the mixed QCD-electroweak non-factorizable 
two-loop contributions to this process. Technically this is the most difficult
part of a complete top-dependent ${\cal O}(g^2 \alpha_s)$ calculation 
because it is the part involving massive two-loop diagrams. 
Of course, for obtaining an infrared 
finite result one needs to add the gluon emission graphs. The remaining 
factorizable and Bremsstrahlung diagrams can be calculated by 
conventional methods.
The ${\cal O}(g^2 \alpha_s)$ gluon emission process involves 
four-point one-loop
diagrams, for which standard techniques such as the Passarino-Veltman reduction
can be used. Here we concentrate only on the pure two-loop part 
of the calculation, for which special massive two-loop methods are necessary.

The coupling of the $Z$ boson to a $b$ quark pair is given by:

\begin{equation}
 - i \frac{g}{2 \cos{\theta_W}} \gamma_{\mu} (v_d - a_d \gamma_5) \;\;\; ,
\end{equation}
which results into a $Z \rightarrow b \bar b$ width given by 
the following expression:

\begin{equation}
  \Gamma_{Z \rightarrow b \bar b} =
  \frac{g^2}{16 \pi \cos^2 \theta_W} M_Z (|v_d|^2 + |a_d|^2) \;\;\; .
\end{equation}
The tree level values
$a_d^{(tree)}  =  - 1/2$
and
$v_d^{(tree)}  =  -1/2 + 2/3 \sin^2 \theta_W$
receive both QCD and electroweak radiative corrections.

The two-loop graphs which contribute to the $Z \rightarrow b\bar{b}$
decay to this order are shown in fig. 1. In the following we 
calculate these graphs for finite Z, W, and top masses, without using
a mass or external momentum expansion. The only approximation we do
in this calculation is to neglect the $b$ quark mass,
where this is justified. We use this
approximation because its effect is very small, 
of the order of $m_b/M_{Z,W,t}$. This approximation simplifies the 
intermediary expressions resulting from the tensor reduction of the diagrams.
However, we note that including a finite $b$ mass throughout our calculation 
would be straightforward.

\begin{figure}
\hspace{1.cm}
    \epsfxsize = 14.5cm
    \epsffile{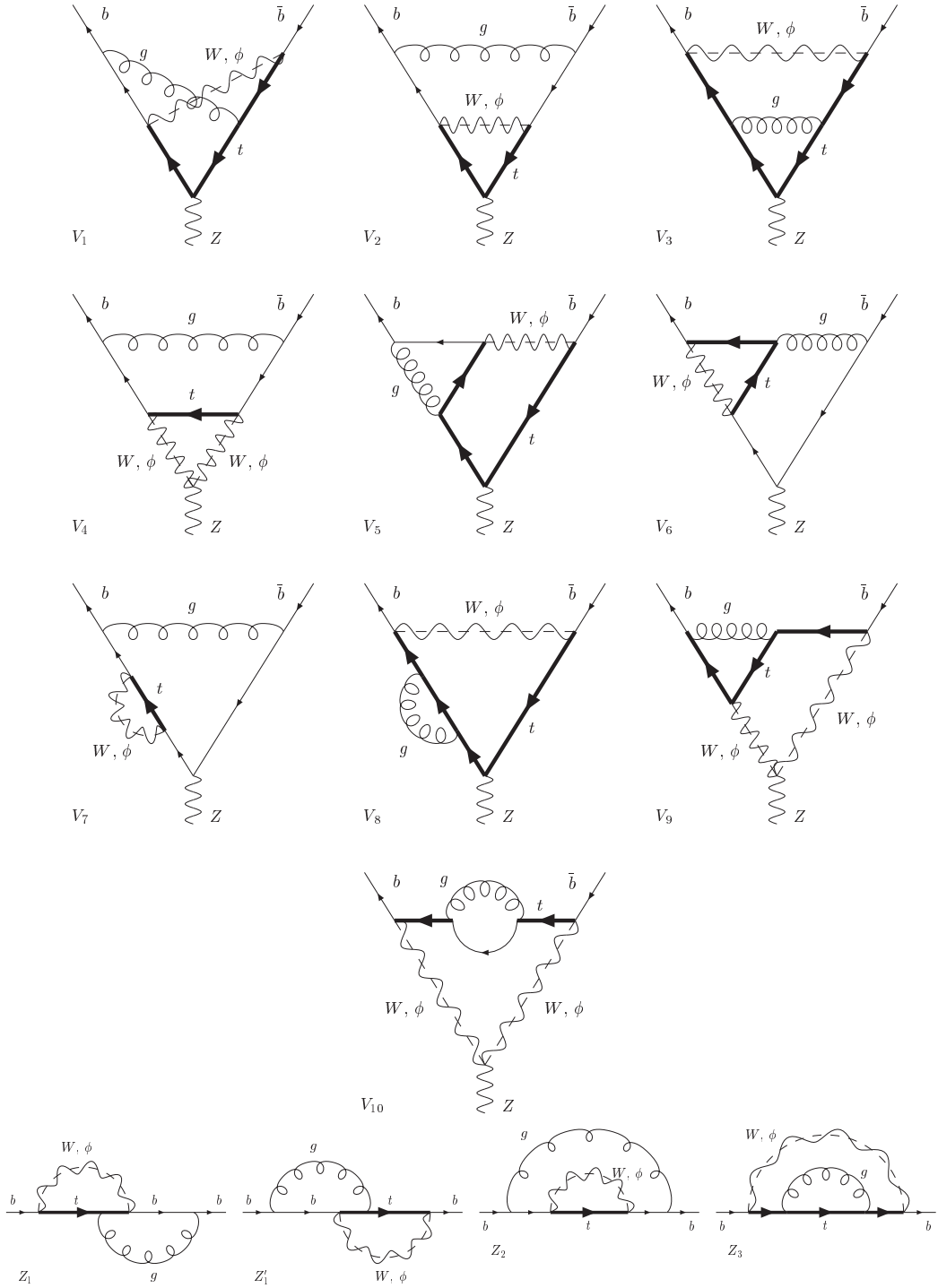}
\caption{{\em The three- and two-point two-loop Feynman graphs which
              contribute to the top-dependent correction 
              of ${\cal O}(\alpha_s g^2)$ to $Z \rightarrow b\bar{b}$.
              Diagrams $V_{1,5,6,7,8,9}$ occur together with the
              diagrams obtained by reverting the fermion line, which
              are not shown explicitly.}}
\end{figure}

The first step in calculating these diagrams, after some trivial 
Dirac algebra, is to decompose the resulting tensor integrals 
into a set of standard scalar integrals. This is done along the lines
of ref. \cite{2loopgeneral}. All tensor structures can be reduced to a set
of ten scalar functions ${\cal H}_1$--${\cal H}_{10}$, whose
definitions are given in ref. \cite{2loopgeneral}. 

In carrying out this scalar decomposition procedure, there is 
a subtlety related to the apparent introduction of some spurious
singularities. To illustrate this issue, let us consider a general 
tensor integral, which is of the following type (see ref. \cite{2loopgeneral}):

\begin{equation}
   \int d^{n}p\,d^{n}q\, 
       \frac{p^{\mu_1} \ldots p^{\mu_i} q^{\mu_{i+1}} \ldots q^{\mu_j}}{
             (p^{2}+m_{1}^{2})^{\alpha_{1}} \,
             (q^{2}+m_{2}^{2})^{\alpha_{2}} \,
             [(r+k)^{2}+m_{3}^{2}]^{\alpha_{3}}
            }
    \; \; .
\end{equation}
In the expression above, all momenta are rotated to Euclidian.
Such a tensor integral can always be decomposed into Lorentz 
invariant integrals of the type:

\begin{equation}
    \int d^{n}p\,d^{n}q\, 
       \frac{(p \cdot k)^a (q \cdot k)^b}{
             (p^{2}+m_{1}^{2})^{\alpha_{1}} \,
             (q^{2}+m_{2}^{2})^{\alpha_{2}} \,
             [(r+k)^{2}+m_{3}^{2}]^{\alpha_{3}}
            }
    \; \; 
\end{equation}
by decomposing the loop integration momenta $p$ and $q$ into components
parallel and orthogonal to the external momentum $k$ of the two-loop
integral:

\begin{equation}
 p_{\perp}^{\mu}  =   p^{\mu} - \frac{p \cdot k}{k^2} k^{\mu} 
 \;\;\; , \;\;\;
 q_{\perp}^{\mu}  =   q^{\mu} - \frac{q \cdot k}{k^2} k^{\mu} 
    \; \; .
\end{equation}

However, in doing this, one apparently introduces light-cone 
singularities of the type $1/k^2$. At the same time, it is obvious 
that the original tensor integral of eq. 3 is free of such singularities,
and in fact it can be evaluated analytically at vanishing external momentum
for any combination of masses \cite{2loopgeneral}. 
This obviously means that the scalar
integrals which result from the tensor decomposition must have such
a $k^2$ behaviour as to compensate the $1/k^2$ singularities 
which result upon introducing the orthogonal loop momenta defined
in eqns. 3. Indeed, one can convince oneself that $1/k^2$ light-cone
singularities are non-existent when the tensor integrals are reduced into 
scalar ${\cal H}_i$ functions.  As an example, we give here the tensor 
decomposition of a two-loop integral with three indices:

\begin{eqnarray}
\lefteqn{   \int d^{n}p\,d^{n}q\, 
       \frac{p^{\mu} q^{\nu} q^{\lambda}}{
    [(p+k)^2+m_1^2]^2 \, (q^2+m_2^2) \, (r^2+m_3^2) }
 \,  =}
 & &
 \nonumber \\
& &               ( \tau^{\mu\nu} k^{\lambda} + \tau^{\mu\lambda} k^{\nu} 
                                     + \tau^{\nu\lambda} k^{\mu} ) 
     \left[ -  \left(\frac{1}{k^2}\right)^2  \frac{n+2}{3(n-1)} {\cal H}_9 \right]
 \nonumber \\
& &              + ( g^{\mu\nu} k^{\lambda} + g^{\mu\lambda} k^{\nu} 
                                     + g^{\nu\lambda} k^{\mu} )    
               \left[ \left(\frac{1}{k^2}\right)^2 \frac{1}{3}  
                   \tilde{{\cal P}}^{12}_{211} \right]
 \nonumber \\ 
 & &              + ( g^{\mu\nu} k^{\lambda} + g^{\mu\lambda} k^{\nu} 
                                  - 2 g^{\nu\lambda} k^{\mu} )    
               \left\{ \frac{1}{k^2} \frac{1}{3}    
              \left[ \tilde{{\cal P}}^{11}_{211} + 
                      \tilde{{\cal P}}^{02}_{211} - \frac{n}{n-1} ({\cal H}_5 + {\cal H}_6) 
             \right] \right\}
  \nonumber 
\end{eqnarray}
where

\begin{equation}
  \tau^{\mu\nu} = g^{\mu\nu} - \frac{k^{\mu}k^{\nu}}{k^2}
\end{equation}
with the scalar functions ${\cal H}$ and $\tilde{{\cal P}}$ defined
in ref. \cite{2loopgeneral}. By using the explicit expressions of the 
${\cal H}$ and $\tilde{{\cal P}}$ functions, one can easily see 
that this decomposition is indeed free of singularities at
$k^2 \rightarrow 0$. Singularity-free decomposition 
formulae are obtained in a similar way for all other tensor 
integrals involved.

We encoded all the necessary Dirac algebra 
into an algebraic manipulation program 
(we used both FORM and Schoonship).
The computer program then uses the two-loop reduction algorithm 
which we described in detail in ref. \cite{2loopgeneral}. Thus it
reduces the two-loop Feynman graphs into a set of standard 
scalar integrals. The final output of the algebraic program
is a combination of the special functions $h_i$, which are the finite
parts of the two-loop scalar integrals ${\cal H}_i$ --- see ref. 
\cite{2loopgeneral}.

These final algebraic expressions are directly suitable for further 
numerical integration. To do this, they are transfered into a FORTRAN
integration package \cite{2loopnumerical}. 
The numerical package can calculate  
efficiently and with high numerical accuracy the functions $h_i$, 
starting from their integral representations. Then it performs the 
numerical integration over the remaining Feynman parameters. For achieving
both a high integration speed and a high accuracy of the final result,
the complex singularities of the integrand are found automatically, and
then a smooth complex integration path is automatically calculated in terms
of spline functions. Then, along this complex path, an adaptative 
deterministic algorithm is used, which leads to an accurate numerical 
evaluation of the Feynman graph.

Collecting all proper vertex contributions shown in figure 1 (diagrams
$V_1$--$V_{10}$), we obtain a correction to the $Zb\bar b$ vertex of the type
$\gamma_{\mu} (1-\gamma_5) A^{(vertex)}$ in the massless $b$ limit. 

From the $b$ self-energy diagrams we derive a $b$ wave function 
renormalization contribution. Denoting the $b$ self-energy by 
$i \Sigma(p)$, we have in general:

\begin{equation}
  \Sigma(p) = \left[ A(p^2) + \gamma_5 A_5(p^2) \right] \gamma \cdot p
              + B(p^2) + \gamma_5 B_5(p^2) \;\; . 
\end{equation}
For the general case of a finite $b$ mass, the coefficients 
$A(p^2)$ and $A_5(p^2)$, and the momentum derivatives of 
$B(p^2)$ and $B_5(p^2)$ at $p^2=m_b^2$ are needed for extracting 
the $b$ wave function renormalization contribution. In ref. \cite{onshellexp}
we have shown how this can be done within our two-loop formalism.
As a side remark, we note here that the on-shell external momentum 
differentiation can be best performed by 
working on the integrand before the momentum integrations 
\cite{onshellexp}. This reduces the size of the expressions which appear
at intermediate stages of the calculation, and provides for a systematic
treatment of all diagrams.
The result for the two-loop wave function renormalization constant 
is again a set of the ten standard functions we have introduced.

In the massless $b$ quark limit, the expressions of the self-energy
diagrams simplify considerably. The self-energy simplifies to 
$\Sigma(p) = A(p^2) (1 + \gamma_5) \gamma \cdot p$. Accordingly, the $b$
wave function renormalization constant contribution to the $Zb\bar b$
vertex correction reads 
$\gamma_{\mu} (1-\gamma_5)(a_d^{tree}+v_d^{(tree)})A(p^2=0)$.

Therefore, the non-factorizable two-loop contributions of figure 1
contribute a correction

\begin{equation}
  \gamma_{\mu} (1-\gamma_5) 
  \left[ A^{(vertex)} + (a_d^{tree}+v_d^{(tree)})A(p^2=0) \right]  
\end{equation}
to the $Zb\bar b$ vertex.

\begin{figure}
\hspace{1.cm}
    \epsfxsize = 14.5cm
    \epsffile{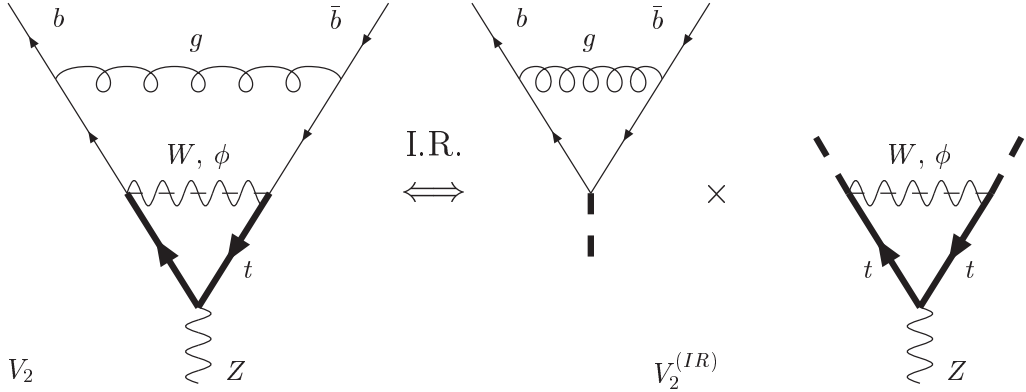}
\caption{{\em Extracting the infrared divergent pieces of the two-loop
              diagrams analytically. The infrared divergency of the two-loop
              diagram is the same as the infrared divergency of the 
              product of the two one-loop diagrams
              obtained by ``freezing'' the common line in the loop momenta 
              integration.}}
\end{figure}

At this point we would like to discuss the treatment of infrared 
divergencies. Among the three-point diagrams, $V_2$, $V_4$, $V_6$,
and $V_7$ are infrared divergent. Their infrared divergencies can be isolated 
analytically in the way shown in figure 2 for the case of $V_2$. 
The analytical
isolation of infrared singularities is based on the observation that 
the original two-loop diagram has one line common for the two loop momenta
which are to be integrated out --- the $W/\phi$ line in the case of 
diagram $V_2$. By ``freezing'' the loop integration momentum of this internal
line to the value of the loop momentum whose one-loop sub-diagram is infrared
finite --- the $ttW$ triangle sub-diagram in the case of diagram $V_2$ --- one
obtains a product of two one-loop diagrams which has precisely the same
infrared singularities as the two-loop diagram. All other
infrared divergent diagrams can be treated in the same way.

The analytical infrared isolation and factorization into one-loop diagrams
clearly works nicely for all two-loop diagrams involved in the 
$Z \rightarrow b \bar b$ calculation, shown in figure 1. However, it is
an open question if this can be done in {\em all possible} two-loop cases,
and especially in the case of calculations of higher order in $\alpha_s$, such
as two-loop pure QCD calculations. This point clearly deserves further 
investigation. We note that our massive two-loop reduction
scheme is mainly designed for {\em massive} calculations; for pure 
QCD calculations other methods which take advantage of the massless 
structure of the theory may prove more efficient --- see for instance 
ref. \cite{qcd} for a few recent examples. 

Once the infrared singularities are isolated and written in the form of 
one-loop diagrams, one chooses a regularization method to treat them and
cancel them upon the real gluon emission process. Dimensional regularization
is often used in higher-order QCD calculations. Given that the process
considered here is only of ${\cal O}(\alpha_s)$ in the strong coupling 
constant, a second possibility is to use a gluon mass regulator. This
does not upset the Slavnov-Taylor identities for our case because to the
order considered here the infrared structure 
is the same as in the Abelian case.

Independently of the choice for the infrared regulator, separating the
infrared singular piece analytically is useful for increasing the 
efficiency of the numerical integration of the two-loop diagram. For instance,
in the case of diagram $V_2$ discussed in figure 2, the infrared singularity
appears as an end-point singularity in a two-fold Feynman parameter 
integration. By subtracting first the product of two one-loop diagrams 
shown in figure 2, this end-point singularity is being removed and the
numerical integration over the resulting smooth function $(V_2-V_2^{(IR)})$ 
becomes much more efficient than $V_2$ alone.

In the following discussion it is understood that the infrared singularities 
are regularized by introducing a gluon mass regulator $m_g$. Again, we would
like to stress that dimensional regularization can be used equally well.

When calculating the two-loop diagrams of figure 1, we deal 
with infrared divergences in the form of $Lim_{m_g\to 0} \ ln(m_g)$, 
where $m_g$ is the gluon mass regulator. These infrared divergent terms
cancel those from real gluon emission  when we calculate a decay rate,
and we have checked this explicitly. The net effect of this cancellation 
is to replace the gluon mass regulator $m_g$ by the maximum undetected gluon 
energy, or the energy resolution of an experiment $\omega_{max}$ --- the 
logarithmic infrared behaviour is universal and we do not need to 
calculate explicitly the Bremsstrahlung process to extract it. 
Of course, after the infrared cancellation one is left with a finite piece
and an explicit calculation of the real gluon emission process is needed.
This involves the evaluation of four-point one-loop
diagrams and integration over three-particle phase space with the 
appropriate experimental cuts. The Bremsstrahlung process is beyond the 
scope of this article --- we only note that the calculation can be 
performed with conventional methods such as the Passarino-Veltman reduction.

\begin{figure}
\hspace{1.cm}
    \epsfxsize = 14.5cm
    \epsffile{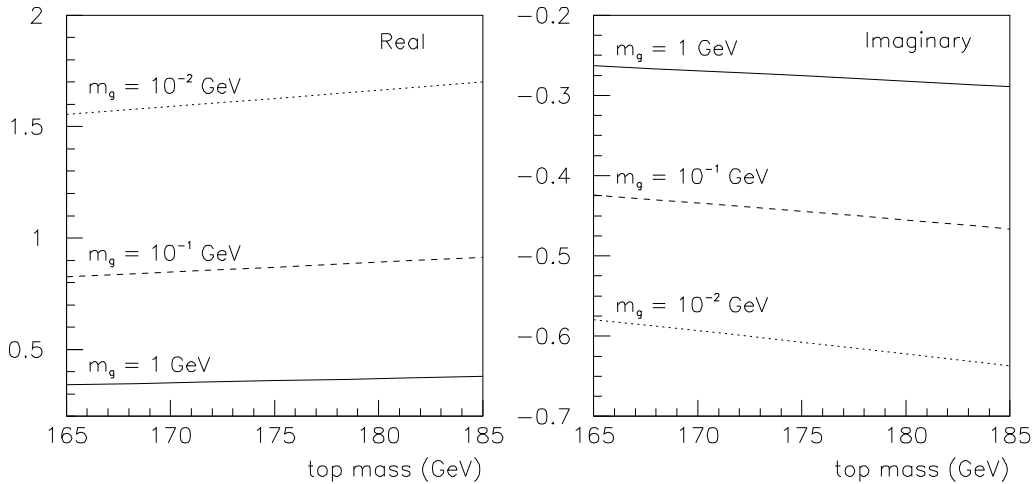}
\caption{{\em Numerical results for the pure two-loop contribution 
              due to the proper Feynman graphs 
              given in figure 1 (see the text). 
              The ultraviolet poles are removed by minimal subtraction.
              The infrared divergencies are separated as one-loop integrals
              and can be handled either by dimensional regularization or
              by using a gluon mass regulator $m_g$. Here we give 
              numerical results for a range of $m_g$. 
              An overall colour and coupling
              constant factor of $i \alpha_s (g^3/12\cos{\theta_W})$
              is understood.}}
\end{figure}

In figure 3 we give  numerical results for the pure two-loop contribution
to the $Z \rightarrow b\bar{b}$ amplitude, stemming from the diagrams shown 
in figure 1. Instead of ploting each diagram separately, we give their
total contribution, which translate into the term
$\left[ A^{(vertex)} + (a_d^{tree}+v_d^{(tree)})A(p^2=0) \right]$ of eq. 8.
All diagrams are subtracted in the ultraviolet by minimal subtraction.
We have checked explicitly that the ultraviolet infinities are 
absorbed into the mass, coupling constant, and wave function renormalization. 
This serves as a good check on the two-loop tensor decomposition algebra.
The infrared divergencies are regulated by using a gluon mass regulator.
Should dimensional regularization of infrared singularities be preferred,
this is straightforward to implement by correspondingly 
treating the one-loop infrared pieces of the type shown in figure 2.
We used 
$M_Z = 91.187$ GeV, $M_W = 80.41$ GeV, 
and the effective electroweak mixing angle
$\sin^2\theta = .2312$.

In conclusion, we calculated all two-loop proper diagrams
relevant for the ${\cal O}(\alpha_s g^2)$ top-dependent correction 
to the $Z \rightarrow b\bar{b}$ decay width, by using two-loop
methods which we developed previously. We have developed an on-shell
momentum expansion which allows for a simple extraction of wave
function renormalization constants. We discussed the isolation of 
infrared singularities in the form of products 
of one-loop integrals. This allows 
the treatment of infrared cancellations with the Bremsstrahlung 
process either by dimensional regularization or by using a gluon 
mass regulator which is legitimate in this particular order. 
Our calculation is exact, 
in the sense that we do not rely upon mass or momentum expansions
of Feynman diagrams --- we rather calculate them at finite mass and
momentum values from the outset. Being based on a general algorithm,
the calculation presented in this contribution opens the way for
the calculation of a number of other two-loop corrections which
have become necessary because of improved experimental electroweak data.

\vspace{1cm}

{\bf Aknowledgements}

The work of A.G. was supported by the US Department of Energy.
The work of Y.-P. Y. was supported partly by the US Department of Energy.



\end{document}